# Superconductive materials with MgB$_2$-like structures from data-driven screening

Ze Yu[1,2], Tao Bo[1,3], Bo Liu[1,2], Zhendong Fu[3], Huan Wang[4,5], Sheng Xu[4,5], Tianlong Xia[4,5]*, Shiliang Li[1,2,3]*, Sheng Meng[1,2,3]*, Miao Liu[1,3,6]*

[1]Beijing National Laboratory for Condensed Matter Physics, Institute of Physics, Chinese Academy of Sciences, Beijing 100190, China

[2]School of Physical Sciences, University of Chinese Academy of Sciences, Beijing 100190, China

[3]Songshan Lake Materials Laboratory, Dongguan, Guangdong 523808, China

[4]Department of Physics, Renmin University of China, Beijing 100872, China

[5]Beijing Key Laboratory of Opto-electronic Functional Materials & Micro-nano Devices, Renmin University of China, Beijing 100872, China

[6]Center of Materials Science and Optoelectronics Engineering, University of Chinese Academy of Sciences, Beijing, 100049, China

*Corresponding author: tlxia@ruc.edu.cn, slli@iphy.ac.cn, smeng@iphy.ac.cn, mliu@iphy.ac.cn

Z. Y. and T. B. contributed equally to this work

## Abstract

Finding viable superconducting materials is of interest to the physics community as the superconductors are the playground to manifest many appealing quantum phenomena. This work exemplifies an end-to-end materials discovery towards novel MgB$_2$-like superconductors, starting from the data-driven compound screening all the way to the experimental materialization. In addition to the known superconducting compounds, CaB$_2$ ($T_c$ = 9.4 ~ 28.6 K), SrGa$_2$ ($T_c$ = 0.1 ~ 2.4 K), BaGa$_2$ ($T_c$ = 0.3 ~ 3.3 K), BaAu$_2$ ($T_c$ = 0.0 ~ 0.5 K), and LaCu$_2$ ($T_c$ = 0.1 ~ 2.2 K) are newly discovered, out of ~182,000 starting structures, to be the most promising superconducting compounds that share similar atomic structures with

MgB$_2$. Moreover, BaGa$_2$ is experimentally synthesized and measured to confirm that the compound is a BCS superconductor with $T_c$ = 0.36 K, in good agreement with our theoretical predictions. This work provides a "once and for all" study for the MgB$_2$-like superconductors and showcases that it is feasible to discover new materials via a data-driven approach.

## Introduction

Superconductivity is a fascinating and robust quantum phenomenon to manifest the many-body correlations in solids. It was first discovered by Heike and Onnes that mercury became superconductive under 4.2 K[1], then various superconducting materials have been found to have elevated $T_c$, including niobium-germanium alloy ($T_c$ = 23.2 K)[2], lanthanum barium copper oxide ($T_c$ = 35.0 K)[3], MgB$_2$ ($T_c$ = 39.0 K)[4], yttrium barium copper oxide ($T_c$ = 92.0 K)[5], pressured H$_3$S ($T_c$ = 203.5 K)[6] and, recently, carbonaceous sulfur hydride under pressure ($T_c$ = 287.7 K)[7]. Among those systems, MgB$_2$ possesses superior properties, warranting its great potential for various applications[8]. For example, the high T$_c$ enables the operation of the MgB$_2$ circuits above 20 K, which is higher than the operating temperature of Nb-based superconducting electronics[9]. In addition, MgB$_2$ can survive under a high magnetic field than that of Nb-based superconductors, therefore MgB$_2$ can be utilized as the magnet material in cryogen-free magnetic resonance imaging (MRI) devices[10-14]. Given the appealing advantages of MgB$_2$, it would be of great interest for the superconductor community to thoroughly screen and evaluate the superconductivity within the MgB$_2$-like structures.

Taking the advantage of recent progress on materials databases, in this work, we obtain 56 compounds in total that share the same MgB$_2$-like crystal structure after the intensive structural screening, which is started from ~18k inorganic compounds from the atomly.net. Further screening based on the thermodynamic stability as well as the dynamic stability narrows down the possibility to only 26 compounds. Since MgB$_2$ was confirmed to satisfy the Bardeen-Cooper-Schrieffer (BCS) theory, the coupling of the electrons to lattice vibrations can be fairly accurately addressed via the first-principles calculations. Hence, all

26 candidates are fed into the following superconducting evaluation at the first-principles level. It is found that the $XB_2$ (X = Mg, Al, Ca, Sc, V, Y, Nb, Ta), $XSi_2$ (X = Ca), and $XGa_2$ (X = Sr, Ba) are superconductive. To validate our approach, one of these compounds, $BaGa_2$, has been experimentally made and tested to confirm that the $BaGa_2$ compound is indeed a superconductor with $T_c$= 0.36 K, in very good agreement with our predictions based on *ab-initio* calculations. This work demonstrates that the BCS superconductors can be discovered in a fairly cost-effective manner through high-throughput computations, and it also provides the superconductor community a viable phase space to search for feasible $MgB_2$-like superconductors.

## Computation details

The materials data used in this work are obtained from the "atomly.net" [15], an open-access density functional theory (DFT) materials data infrastructure which is conceptually similar to the Materials Project[16], AFLOWlib[17], and OQMD[18]. The data are generated through high-throughput first-principles calculations with an automated workflow to crunch through more than 180,000 inorganic crystal materials. The dataset contains the crystal structure (*e.g.*, crystal structure, X-ray diffraction (XRD)), electronic structure (*e.g.*, the density of states (DOS) and bandgap ($E_g$)) and energies (*e.g.*, formation energy ($E_{form}$), energy above the convex hull ($E_{hull}$)) of the compounds, basically all the properties which can be obtained from first-principles calculation. The data are calculated by using the "Vienna Ab Initio Simulation Package" (VASP)[19], and the Perdew-Burke-Ernzerhof (PBE)[20] exchange-correlation functional of generalized gradient approximation (GGA) to deal with the interactions between electrons. The cutoff energy for plane waves is 520 eV, and the k-mesh is 3000/Å$^3$ or higher.

After the screening, structural optimization is performed for all candidate materials. The plane-wave cutoff was set to be 520 eV with a k-point mesh of (17 × 17 × 15) in the Monkhorst-Pack sampling scheme. For geometric optimization, all atoms are allowed to fully relax until the forces on atoms are less than 0.01 eV/Å, after considering spin polarization.

The phonon dispersion, electron-phonon coupling (EPC), and superconducting properties are performed using QUANTUM-ESPRESSO (QE) package[21]. The plane-wave kinetic-energy cutoff and the energy cutoff for charge density are set as 100 and 800 Ry, respectively. The Brillouin zone (BZ) k-point mesh of 24 × 24 × 24 and a Methfessel-Paxton smearing width of 0.02 Ry are used to calculate the self-consistent electron density. The dynamic matrix and EPC matrix elements are computed within an 8 × 8 × 8 q mesh. The phonon properties and EPC are calculated based on the density functional perturbation theory (DFPT)[22] and Eliashberg theory[23]. The calculation details are shown in the supplementary materials[24].

Single crystals of $BaGa_2$ were grown by the self-flux method as reported previously[25]. The resistivity, magnetic susceptibility, and specific heat were measured on a physical property measurement system (PPMS, Quantum Design) with the dilution refrigerator insert.

## Results and discussions

In order to comprehensively screen the $MgB_2$-like materials from the database, the $MgB_2$ structure (space group P6/mmm, ICSD ID 193379, Materials Project ID mp-763, and atomly ID 0000105522) is used as the structural template. There are two planar layers of atoms in each primitive cell, and the metallic element (X) forms a triangle lattice layer and the non-metallic element (Y) forms the honeycomb lattice layer, and the two planar layers repeat themselves in *c* direction alternatively, forming the stoichiometry of $XY_2$, as shown in **Fig. 1a**.

A screening process, as shown in **Fig. 1b**, is performed on a dataset of the 182155 materials structures, which is obtained from the atomly.net database, to filter out all the compounds with $MgB_2$ geometry. An in-house script along with the StructureMatcher module of the Pymatgen library[26] is employed to extract all the compounds with the $MgB_2$-like structure, and all the radioactive-element-containing compounds are excluded. After the rigorous filtration process, 56 compounds fall out to have the $MgB_2$-like structure. Then the following screening on the thermodynamic stability narrows down the candidate materials to 51 compounds. The thermodynamic stability of the compounds is assessed by the physical quantity of energy above hull ($E_{hull}$), which is the reaction enthalpy required to decompose a material to other stable compounds. The detailed definition of the $E_{hull}$ can be found in ref. [27,28], and thermodynamics-wise, compound with $E_{hull}$ = 0 meV/atom is the most stable compound, and the stability is worsening when $E_{hull}$ increases. Out of the 56 candidates, the $E_{hull}$ of 51 structures are smaller than 200 meV/atom, 43 structures are smaller than 100 meV/atom, and 35 structures are smaller than 50 meV/atom. Normally, the threshold of 200 meV/atom is a good value to indicate that the compound is likely to be synthesized, and the 50 meV/atom means the compound is highly likely to be synthesized[29,30]. Finally, the dynamic stability assessment based on phonon spectrum calculations shows that there are 26 dynamically stable materials. They are $MgB_2$, $AlB_2$, $CaB_2$, $ScB_2$, $TiB_2$, $VB_2$, $YB_2$, $ZrB_2$, $NbB_2$, $SmB_2$, $GdB_2$, $TbB_2$, $DyB_2$, $HoB_2$, $ErB_2$, $TmB_2$, $HfB_2$, $TaB_2$, $CaSi_2$, $GdSi_2$, $ErSi_2$, $TmSi_2$, $SrGa_2$, $BaGa_2$, $BaAu_2$, and $LaCu_2$ (see **Fig. S1-4** in the Supplemental Material[24]). Other compounds, $BeB_2$, $CrB_2$, etc., are found to be dynamically unstable due to the presence of imaginary frequencies in their phonon spectrum. According to the species of anions, these 26 materials can be divided into four categories: the borides $XB_2$ (X = Mg, Al, Ca, Sc, Ti, V, Y, Zr, Nb, Sm, Gd, Tb, Dy, Ho, Er, Tm, Hf, Ta), the silicides $XSi_2$ (X = Ca, Gd, Er, Tm), the gallides $XGa_2$ (X = Sr, Ba) and the alloys ($BaAu_2$, $LaCu_2$).

According to the existing literature, some of these materials have been extensively studied previously[31-41]. For example, $MgB_2$ and $NbB_2$, have been found to transit to superconducting state below 39.0 K and 9.2 K, respectively. Since $MgB_2$ belongs to the category of conventional superconductors (BCS superconductors), where BCS theory applies, the electron-phonon coupling, as well as the superconducting transition temperature $T_c$, can be obtained from first-principles calculations.

For the 26 dynamic stable candidates, the EPC constant λ and superconducting transition temperature $T_c$ are obtained from DFPT calculations, as summarized in **Table I**. $MgB_2$ has the highest calculated $T_c$ of 37.9 K when the empirical parameter $\mu^*$ that represents the effective screened-Coulomb-repulsion is 0.04, which is in good agreement with the $T_c$ = 39 K obtained in the experiment while tolerating computation errors. Previously, it was found in experiments that VB$_2$[42], TiB$_2$[42], and HfB$_2$[42] are non-superconductive, in good agreement with our prediction. It was also found that ZrB$_2$[31,42], NbB$_2$[42-48], and TaB$_2$[35,42,49] are superconductive with $T_c$ up to 5.5 K, 9.2 K, and 9.5 K, respectively, confirming our predictions too. In those experiments, the $T_c$ is highly tunable by *non-stoichiometric* ratio, external impurities, etc, hence the experimental values of $T_c$ may still need further investigation. CaSi$_2$ has a $T_c$ of 14 K when it is under high pressure[50], in line with our predicted value, which is $T_c$ = 8.7 ~ 16.0 K. The rare-earth-containing compounds, especially lanthanides (SmB$_2$, GdB$_2$, TbB$_2$, DyB$_2$, HoB$_2$, ErB$_2$, TmB$_2$, HfB$_2$, GdSi$_2$, ErSi$_2$, TmSi$_2$), are not superconductors based on our *ab-initio* calculations, confirming the experimental data[51]. AlB$_2$ and YB$_2$ are two exceptions as they are superconductive according to our calculation, but non-superconductive from real-world experiments, therefore they worth in-depth studies.

Other than the compounds mentioned above, the present work adds CaB$_2$, SrGa$_2$, BaGa$_2$, BaAu$_2$, and LaCu$_2$ to the list of compounds within the MgB$_2$-structural family, as they have not yet been experimentally examined. Based on our theoretical evaluation, CaB$_2$, SrGa$_2$, BaGa$_2$, BaAu$_2$, and LaCu$_2$ are highly likely to be superconductive, whereas the GdSi$_2$, ErSi$_2$, and TmSi$_2$ are non-superconductive. The $T_c$ for CaB$_2$, SrGa$_2$, BaGa$_2$, BaAu$_2$, and LaCu$_2$, are 9.4 ~ 28.6 K, 0.1 ~ 2.4 K, 0.3 ~ 3.3 K, 0.0 ~ 0.5 K, 0.1 ~ 2.2 K, respectively. Recently, a theoretical work also found that the superconducting transition temperature $T_c$ of BaGa$_2$ is 1.2 K[52].

To investigate the EPC and superconducting transition temperature $T_c$ of CaB$_2$, SrGa$_2$, and BaGa$_2$, we calculate their phonon dispersion, the phononic density of states (PhDOS), Eliashberg spectral function $\alpha^2F(\omega)$ and cumulative frequency-dependent of EPC constant $\lambda(\omega)$ as presented in **Fig. 2**. For CaB$_2$, the Ca vibrations mainly dominate the low frequencies of 0-300 cm$^{-1}$, while the B vibrations mainly dominate the high frequencies of above 300 cm$^{-1}$ (see **Fig. 2a** and **c**). There are some softened phonon modes along the Γ-A line and around the Γ point at about 350 cm$^{-1}$, which are associated with the B$_{xy}$ vibration, yielding a large EPC $\lambda_{qv}$, as shown in **Fig. 2b**. The phonon dispersion in the frequency range of 300-600 cm$^{-1}$, contributes most to the EPC $\lambda_{qv}$ (see **Fig. 2d**). The calculated EPC $\lambda(\omega)$ of CaB$_2$ is 0.75, and when $\mu^*$ is 0.14 (the same as for SrGa$_2$ and BaGa$_2$), the corresponding $T_c$ is 14.0 K which is quite high in this series of materials.

In SrGa$_2$ and BaGa$_2$, some physical properties such as the vibration dominance are quite different. For SrGa$_2$, the Sr vibrations mainly dominate the low frequencies of 0-120 cm$^{-1}$, while the Ga vibrations spread in the whole area of the Brillouin zone (BZ) (see **Fig. 2e and f**). There is a softened phonon mode around the A point, associated with the Ga$_z$ vibration (~40 cm$^{-1}$), yielding a large EPC $\lambda_{qv}$, as shown in **Fig. 2f**. In addition, the phonon dispersion in the frequency range of 40-120 cm$^{-1}$, contributes most to the EPC $\lambda_{qv}$. The calculated EPC $\lambda(\omega)$ and $T_c$ for SrGa$_2$ are 0.44 and 0.4 K, respectively. The low-frequency phonons (40-120 cm$^{-1}$) contribute 86% of the EPC $\lambda$, while the high-frequency phonons (180-220 cm$^{-1}$) contribute about 14% (see **Fig. 2h**). The EPC properties of BaGa$_2$ are roughly the same as SrGa$_2$, but relatively stronger. As shown in **Fig. 2i** and **j**, the Ba vibrations mainly dominate the low frequencies of 0-100 cm$^{-1}$, while the Ga vibrations spread in the whole area of the Brillouin zone (BZ). No imaginary phonon modes indicate that it is dynamically stable, at least from the perspective of the *ab-initio* calculations. As shown in **Fig. 2j and l**, the low-frequency phonons (40-100 cm$^{-1}$) have a large contribution to the EPC $\lambda_{qv}$ and total EPC $\lambda$. The calculated EPC $\lambda(\omega)$ and $T_c$ are 0.56 and 1.3 K, a little lower than those in CaB$_2$ and SrGa$_2$.

To validate the high-throughput data-driven prediction based on first principles, the BaGa$_2$ compound is experimentally synthesized and characterized for its superconductivity. We found that the BaGa$_2$ is indeed a superconductor with a measured $T_c$ = 0.36 K, in good agreement with the theoretical prediction, which is 0.3 ~ 3.3 K. The superconducting properties were measured by resistivity, magnetic-susceptibility, and heat-capacity measurements. **Fig. 3a** shows the temperature dependence of the resistance, which gives the onset $T_c$ of about 0.37 K. Zero resistance is found below 0.29 K. **Fig. 3b** shows the real component of the AC susceptibility $\chi'$ as the function of temperature, which exhibits Meissner effect below 0.32 K. **Fig. 3c** further shows the temperature dependence of the specific heat. Above 0.4 K, the specific heat is linearly dependent on the temperature, demonstrating that the phonon contribution can be neglected below 1 K. A superconducting jump is found at 0.34 K and disappears under 5 kOe. The behavior below $T_c$ can be well fitted by the BCS expression for the specific heat,

$$C = \frac{T}{T_c}\frac{d}{dt}\int_0^\infty dy \left(-\frac{6\gamma\Delta_0}{k_B\pi}\right)[f\ln f + (1-f)\ln(1-f)]$$
(1)

where $\gamma$ is the normal-state Sommerfeld coefficient and $f$ is the Fermi-Dirac distribution function, $f = 1/(e^{E/k_BT} + 1)$. The energy of quasiparticles is given by $E = \sqrt{\epsilon^2 + \Delta^2}$, where $\epsilon$ is the energy of the normal electrons relative to Fermi surface. The integration variable is $y = \epsilon/\Delta$. Unlike MgB$_2$[53], there is no need to introduce two gaps to describe the specific heat data. The fitted value of $\Delta_0$, the superconducting gap at 0 K, is 0.0433 meV, which gives $2\Delta_0/k_BT_c = 2.83$ with $T_c$ = 0.355 K.

The CaB$_2$, which is predicted to have a higher $T_c$, has a moderate thermal stability as the $E_{hull}$ is 191 meV/atom, hence, it is harder to be synthesized via the traditional route. Sometimes soft chemistry, template epitaxial growth, or high-pressure synthesis may be able to overcome the unfavorable phase competition. Combining all the information obtained here, SrGa$_2$, BaAu$_2$, LaCu$_2$, and CaB$_2$, are worth further experimental explorations.

## Conclusion

The vast phase space of the MgB$_2$-like is thoroughly explored to discover new superconductive compounds, leveraging the recent advances of high-throughput computing, materials databases, first-principles calculations as well as state-of-art experiments. This work finds that the SrGa$_2$, BaGa$_2$, BaAu$_2$, and LaCu$_2$ are promising superconductors, falling out of ~182 thousand starting structures, and of which the BaGa$_2$ is experimentally synthesized and measured to confirm that the compound is a BCS superconductor with $T_c$ = 0.36 K, in line with theoretical predictions. The CaB$_2$ is predicted to have a higher $T_c$, but the synthesis can be very challenging. This work provides useful hints for the superconductor community to search for feasible MgB$_2$-like BCS superconductors with good performance. From a broad materials science point of view, this work demonstrates an end-to-end materials discovery strategy to forcefully filter out the desired materials via a fast and cost-effective fashion. The "fourth paradigm" data-intensive scientific discovery of promising superconductors is on the radar.

## Acknowledgments

The computational resource is provided by the Platform for Data-Driven Computational Materials Discovery of the Songshan Lake laboratory. We especially thank the Atomly database for data sharing. We would also acknowledge the financial support from the Chinese Academy of Sciences (No. ZDBS-LY-SLH007, No. XDB33020000, CAS-WX2021PY-0102), National Science Foundation of China (12025407 and 11934003), and Ministry of Science and Technology (2021YFA1400200).

## Supplementary Information

See the Supplementary Information.

## Figure captions

**Figure 1**. (a) The side & top view of the MgB$_2$-like materials. (b) The screening flow chart for MgB$_2$-like materials. The total process is divided into three steps: filtering out all MgB$_2$-like materials, evaluating thermodynamic stability and dynamics stability. The thermodynamically stable materials are summarized in the table on the right, while the dynamically stable ones are summarized in the table below.

**Figure 2**. The phonon dispersion weighted by motion modes of constituent elements, the phonon dispersion weighted by the magnitude of EPC $\lambda_{qv}$, PhDOS, Eliashberg spectral function $\alpha^2F(\omega)$ and cumulative frequency-dependent of EPC constant $\lambda(\omega)$ of (a-d) CaB$_2$, (e-h) SrGa$_2$, and (i-l) BaGa$_2$. The red, green, blue, and orange colors in (a & e & i) represent B$_{xy}$(Ga$_{xy}$, Ga$_{xy}$), B$_z$(Ga$_z$, Ga$_z$), Ca$_{xy}$(Sr$_{xy}$, Ba$_{xy}$), and Ca$_z$(Sr$_z$, Ba$_z$) modes, respectively.

**Figure 3** (a) Temperature dependence of the resistivity. Due to the very small resistance, a current of 0.3 mA was used. (b) Temperature dependence of the real component of the AC susceptibility. (c) Temperature dependence of the specific heat at various field. The solid line is fitted by Eq. (1).

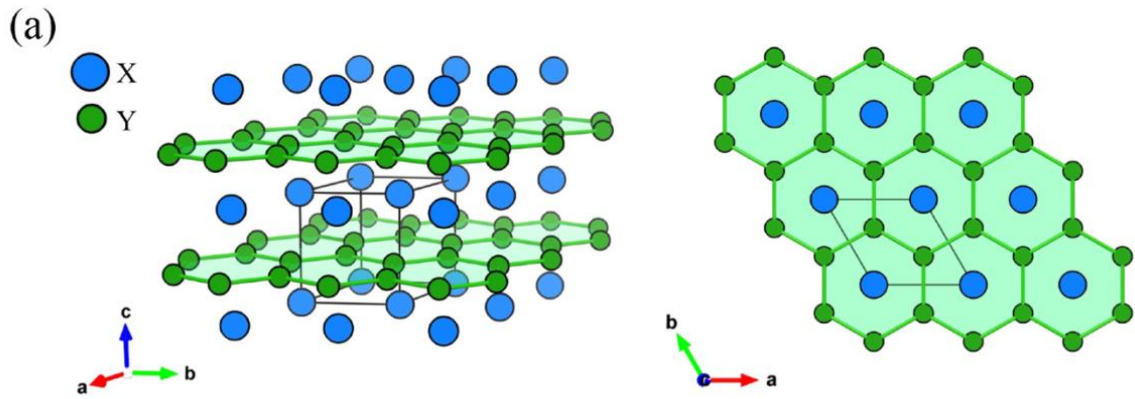

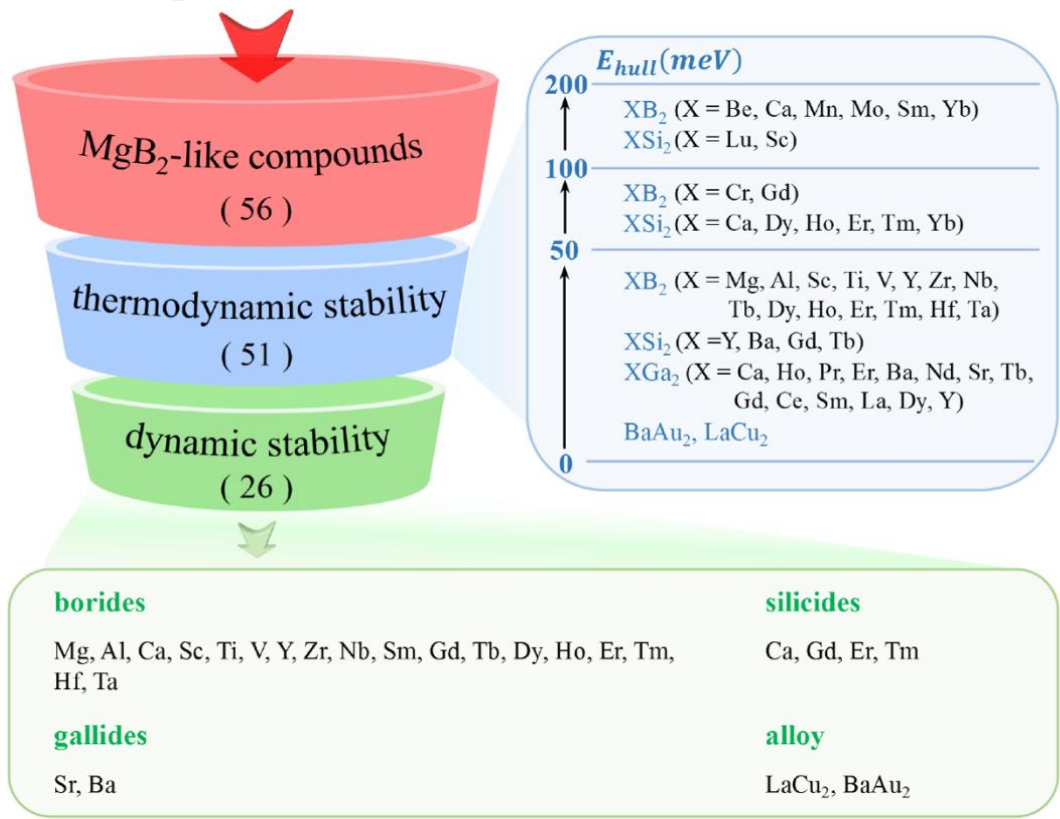

**Figure. 1**

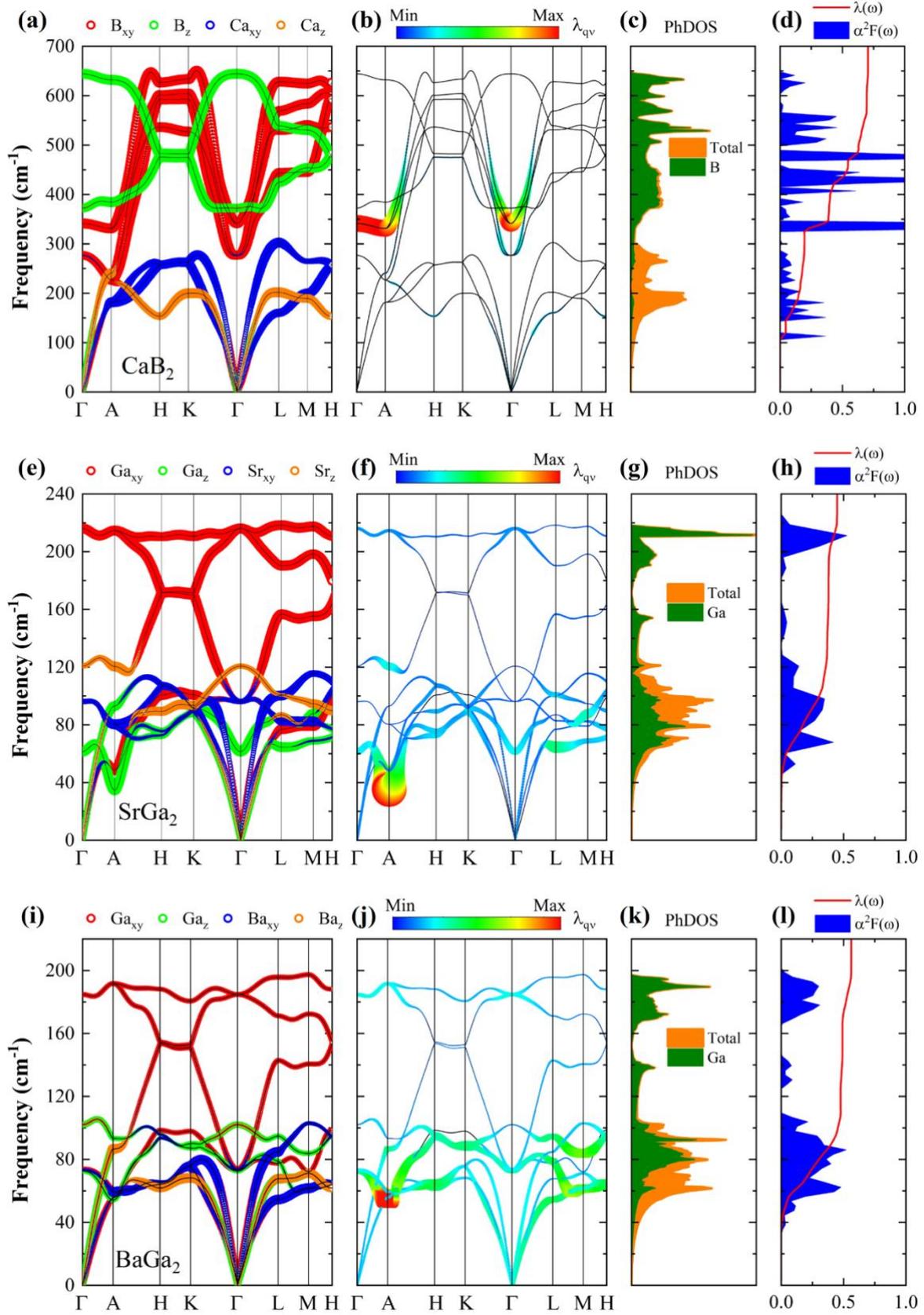

Figure. 2

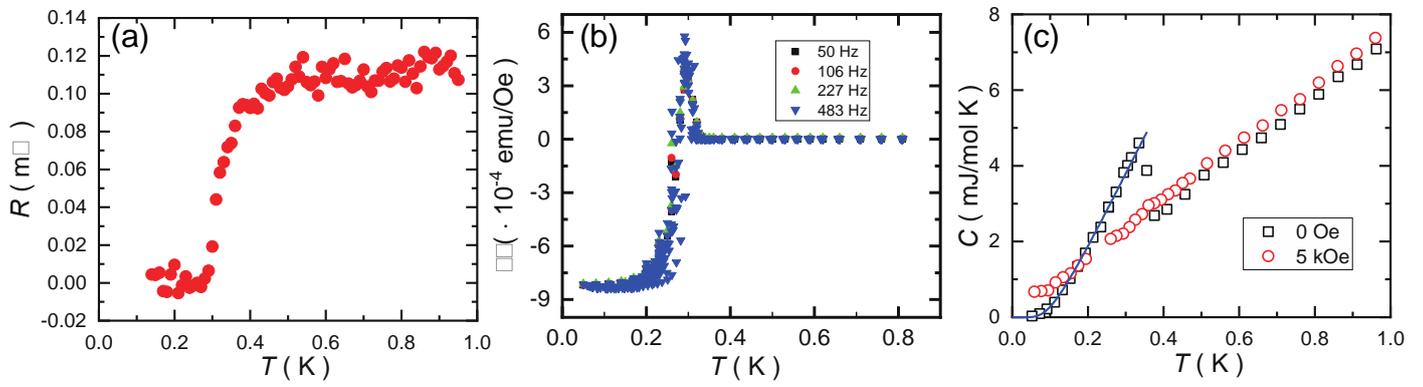

**Figure. 3**

**Table I**. The total EPC λ, $T_c^{cal}$ calculated by the *ab-initio* calculations, $T_c^{exp}$ measured by experiments and corresponding references. The empirical parameter $\mu^*$ representing effective shielding of Coulomb rejection ranges from 0.04 to 0.18. Symbol notes: "/" means or, "~" means within this range, "-" means no corresponding data.

| Formula | λ | $T_c^{cal}$(K) | $T_c^{exp}$(K) | Reference |
|---|---|---|---|---|
| **MgB$_2$** | 0.68 | 10.3 ~ 37.9 | 39.0 | [4] |
| **AlB$_2$** | 0.47 | 0.6 ~ 10.9 | 0.0 | [54] |
| **CaB$_2$** | 0.75 | 9.4 ~ 28.6 | - | - |
| **ScB$_2$** | 0.28 | 0.0 ~ 1.7 | 1.5 | [55,56] |
| **TiB$_2$** | 0.13 | 0.0 ~ 0.0 | 0.0 | [42] |
| **VB$_2$** | 0.36 | 0.0 ~ 4.2 | 0.0 | [42] |
| **YB$_2$** | 0.40 | 0.1 ~ 6.1 | 0.0 | [57] |
| **ZrB$_2$** | 0.14 | 0.0 ~ 0.0 | 0.0 / 5.5 | [31,42] |
| **NbB$_2$** | 0.71 | 5.7 ~ 19.4 | 0.6 ~ 9.2 | [42-48] |
| **SmB$_2$** | 0.08 | 0.0 ~ 0.0 | 0.0 | [51] |
| **GdB$_2$** | 0.15 | 0.0 ~ 0.0 | 0.0 | [51] |
| **TbB$_2$** | 0.12 | 0.0 ~ 0.0 | 0.0 | [51] |
| **DyB$_2$** | 0.11 | 0.0 ~ 0.0 | 0.0 | [51] |
| **HoB$_2$** | 0.10 | 0.0 ~ 0.0 | 0.0 | [51] |
| **ErB$_2$** | 0.09 | 0.0 ~ 0.0 | 0.0 | [51] |
| **TmB$_2$** | 0.10 | 0.0 ~ 0.0 | 0.0 | [51] |
| **HfB$_2$** | 0.14 | 0.0 ~ 0.0 | 0.0 | [42] |
| **TaB$_2$** | 0.95 | 9.4 ~ 20.3 | 0.0 / 9.5 | [35,42,49] |
| **CaSi$_2$** | 1.13 | 8.7 ~ 16.0 | 14.0 | [50] |
| **GdSi$_2$** | 0.07 | 0.0 ~ 0.0 | - | - |
| **ErSi$_2$** | 0.09 | 0.0 ~ 0.0 | - | - |
| **TmSi$_2$** | 0.12 | 0.0 ~ 0.0 | - | - |
| **SrGa$_2$** | 0.44 | 0.1 ~ 2.4 | - | - |
| **BaGa$_2$** | 0.56 | 0.3 ~ 3.3 | - | - |
| **BaAu$_2$** | 0.34 | 0.0 ~ 0.5 | - | - |
| **LaCu$_2$** | 0.46 | 0.1 ~ 2.2 | - | - |

## References

[1] H. K. Onnes, in *KNAW, Proceedings* 1911), pp. 1910.
[2] J. R. Gavaler, Superconductivity in Nb–Ge films above 22 K, Appl. Phys. Lett. **23**, 480 (1973).
[3] J. G. Bednorz and K. A. MüLler, Possible high Tc superconductivity in the Ba-La-Cu-O system, Zeitschrift für Physik B Condensed Matter **64**, 189 (1986).
[4] J. Nagamatsu, N. Nakagawa, T. Muranaka, Y. Zenitani, and J. Akimitsu, Superconductivity at 39 K in magnesium diboride, Nature **410**, 63 (2001).
[5] M. K. Wu, J. R. Ashburn, C. J. Torng, P. H. Hor, R. L. Meng, L. Gao, Z. J. Huang, Y. Q. Wang, and C. W. Chu, Superconductivity at 93 K in a new mixed-phase Y-Ba-Cu-O compound system at ambient pressure, Phys. Rev. Lett. **58**, 908 (1987).
[6] A. P. Drozdov, M. I. Eremets, I. A. Troyan, V. Ksenofontov, and S. I. Shylin, Conventional superconductivity at 203 kelvin at high pressures in the sulfur hydride system, Nature **525**, 73 (2015).
[7] E. Snider, N. Dasenbrock-Gammon, R. Mcbride, M. Debessai, H. Vindana, K. Vencatasamy, K. V. Lawler, A. Salamat, and R. P. Dias, Room-temperature superconductivity in a carbonaceous sulfur hydride, Nature **586**, 373 (2020).
[8] M. Tomsic, M. Rindfleisch, J. Yue, K. Mcfadden, J. Phillips, M. D. Sumption, M. Bhatia, S. Bohnenstiehl, and E. W. Collings, Overview of $MgB_2$ Superconductor Applications, International Journal of Applied Ceramic Technology **4**, 250 (2007).
[9] D. Cunnane, K. Chen, and X. X. Xi, Superconducting $MgB_2$ rapid single flux quantum toggle flip flop circuit, Appl. Phys. Lett. **102**, 222601 (2013).
[10] M. Modica, S. Angius, L. Bertora, D. Damiani, M. Marabotto, D. Nardelli, M. Perrella, M. Razeti, and M. Tassisto, Design, construction and tests of $MgB_2$ coils for the development of a cryogen free magnet, IEEE transactions on applied superconductivity **17**, 2196 (2007).
[11] C. Poole, T. Baig, R. J. Deissler, D. Doll, M. Tomsic, and M. Martens, Numerical study on the quench propagation in a 1.5 T $MgB_2$ MRI magnet design with varied wire compositions, Supercon. Sci. Technol. **29**, 044003 (2016).
[12] J. Ling, J. P. Voccio, S. Hahn, T. Qu, J. Bascuñán, and Y. Iwasa, A persistent-mode 0.5 T solid-nitrogen-cooled $MgB_2$ magnet for MRI, Supercon. Sci. Technol. **30**, 024011 (2016).
[13] T. Baig, A. Al Amin, R. J. Deissler, L. Sabri, C. Poole, R. W. Brown, M. Tomsic, D. Doll, M. Rindfleisch, and X. Peng, Conceptual designs of conduction cooled $MgB_2$ magnets for 1.5 and 3.0 T full body MRI systems, Supercon. Sci. Technol. **30**, 043002 (2017).
[14] D. Patel, A. Matsumoto, H. Kumakura, M. Maeda, S.-H. Kim, H. Liang, Y. Yamauchi, S. Choi, J. H. Kim, and M. S. A. Hossain, Superconducting joints using multifilament $MgB_2$ wires for MRI application, Scripta Materialia **204**, 114156 (2021).
[15] Atomly, https://atomly.net.
[16] A. Jain, S. P. Ong, G. Hautier, W. Chen, W. D. Richards, S. Dacek, S. Cholia, D. Gunter, D. Skinner, G. Ceder, and K. A. Persson, Commentary: The Materials Project: A materials genome approach to accelerating materials innovation, APL Materials **1**, 011002 (2013).
[17] S. Curtarolo, W. Setyawan, G. L. W. Hart, M. Jahnatek, R. V. Chepulskii, R. H. Taylor, S. Wang, J. Xue, K. Yang, O. Levy, M. J. Mehl, H. T. Stokes, D. O. Demchenko, and D. Morgan, AFLOW: An automatic framework for high-throughput materials discovery, Computational Materials Science **58**, 218 (2012).
[18] J. E. Saal, S. Kirklin, M. Aykol, B. Meredig, and C. Wolverton, Materials Design and Discovery with High-Throughput Density Functional Theory: The Open Quantum Materials Database (OQMD), JOM **65**, 1501 (2013).
[19] G. Kresse and J. Furthmüller, Efficient iterative schemes for ab initio total-energy calculations using a plane-wave basis set, Phys. Rev. B **54**, 11169 (1996).
[20] J. P. Perdew, K. Burke, and M. Ernzerhof, Generalized Gradient Approximation Made Simple, Phys. Rev. Lett. **77**, 3865 (1996).
[21] P. Giannozzi, S. Baroni, N. Bonini, M. Calandra, R. Car, C. Cavazzoni, D. Ceresoli, G. L. Chiarotti, M. Cococcioni, and I. Dabo, QUANTUM ESPRESSO: a modular and open-source software project for quantum simulations of materials, Journal of physics: Condensed matter **21**, 395502 (2009).
[22] S. Baroni, S. De Gironcoli, A. Dal Corso, and P. Giannozzi, Phonons and related crystal properties from density-functional perturbation theory, Rev. Mod. Phys. **73**, 515 (2001).

[23] F. Giustino, Electron-phonon interactions from first principles, Rev. Mod. Phys. **89** (2017).
[24] See Supplemental Material at [URL will be inserted by publisher] for Phonon spectrums.
[25] S. Xu, C. Bao, P.-J. Guo, Y.-Y. Wang, Q.-H. Yu, L.-L. Sun, Y. Su, K. Liu, Z.-Y. Lu, S. Zhou, and T.-L. Xia, Interlayer quantum transport in Dirac semimetal BaGa2, Nat Commun **11** (2020).
[26] S. P. Ong, W. D. Richards, A. Jain, G. Hautier, M. Kocher, S. Cholia, D. Gunter, V. L. Chevrier, K. A. Persson, and G. Ceder, Python Materials Genomics (pymatgen): A robust, open-source python library for materials analysis, Computational Materials Science **68**, 314 (2013).
[27] S. P. Ong, L. Wang, B. Kang, and G. Ceder, Li−Fe−P−O2 Phase Diagram from First Principles Calculations, Chemistry of Materials **20**, 1798 (2008).
[28] A. Jain, G. Hautier, S. P. Ong, C. J. Moore, C. C. Fischer, K. A. Persson, and G. Ceder, Formation enthalpies by mixing GGA and GGA+Ucalculations, Phys. Rev. B **84** (2011).
[29] M. Liu, A. Jain, Z. Rong, X. Qu, P. Canepa, R. Malik, G. Ceder, and K. A. Persson, Evaluation of sulfur spinel compounds for multivalent battery cathode applications, Energy & Environmental Science **9**, 3201 (2016).
[30] M. Aykol, S. S. Dwaraknath, W. Sun, and K. A. Persson, Thermodynamic limit for synthesis of metastable inorganic materials, Science advances **4**, eaaq0148 (2018).
[31] V. A. Gasparov, N. S. Sidorov, I. I. Zver'Kova, and M. P. Kulakov, Electron transport in diborides: Observation of superconductivity in ZrB2, Journal of Experimental and Theoretical Physics Letters **73**, 532 (2001).
[32] D. Kaczorowski, J. Klamut, and A. Zaleski, Some comments on superconductivity in diborides, arXiv preprint cond-mat/0104479 (2001).
[33] N. I. Medvedeva, A. L. Ivanovskii, J. E. Medvedeva, and A. J. Freeman, Electronic structure of superconductingMgB2and related binary and ternary borides, Phys. Rev. B **64** (2001).
[34] P. Ravindran, P. Vajeeston, R. Vidya, A. Kjekshus, and H. Fjellvåg, Detailed electronic structure studies on superconductingMgB2and related compounds, Phys. Rev. B **64** (2001).
[35] H. Rosner, W. E. Pickett, S.-L. Drechsler, A. Handstein, G. Behr, G. Fuchs, K. Nenkov, K.-H. Müller, and H. Eschrig, Electronic structure and weak electron-phonon coupling inTaB2, Phys. Rev. B **64** (2001).
[36] G. Satta, G. Profeta, F. Bernardini, A. Continenza, and S. Massidda, Electronic and structural properties of superconductingMgB2,CaSi2,and related compounds, Phys. Rev. B **64** (2001).
[37] S. Elgazzar, P. M. Oppeneer, S.-L. Drechsler, R. Hayn, and H. Rosner, Calculated de Haas-van Alphen data and plasma frequencies of MgB2 and TaB2, Solid State Commun. **121**, 99 (2002).
[38] T. Takahashi, S. Kawamata, S. Noguchi, and T. Ishida, Superconductivity and crystal growth of NbB2, Physica C: Superconductivity **426-431**, 478 (2005).
[39] H. J. Choi, S. G. Louie, and M. L. Cohen, Prediction of superconducting properties ofCaB2using anisotropic Eliashberg theory, Phys. Rev. B **80** (2009).
[40] S. Okatov, A. Ivanovskii, Y. E. Medvedeva, and N. Medvedeva, The electronic band structures of superconducting MgB2 and related borides CaB2, MgB6 and CaB6, physica status solidi (b) **225**, R3 (2001).
[41] T. Oguchi, Cohesion in AlB2-type diborides: a first-principles study, J. Phys. Soc. Jpn. **71**, 1495 (2002).
[42] L. Leyarovska and E. Leyarovski, A search for superconductivity below 1 k in transition metal borides, Journal of the Less Common Metals **67**, 249 (1979).
[43] A. Yamamoto, C. Takao, T. Masui, M. Izumi, and S. Tajima, High-pressure synthesis of superconducting Nb1−xB2 (x=0–0.48) with the maximum Tc=9.2 K, Physica C: Superconductivity **383**, 197 (2002).
[44] H. Takeya, K. Togano, Y. S. Sung, T. Mochiku, and K. Hirata, Metastable superconductivity in niobium diborides, Physica C: Superconductivity **408**, 144 (2004).
[45] J. K. Hulm and B. T. Matthias, New Superconducting Borides and Nitrides, Phys. Rev. **82**, 273 (1951).
[46] W. T. Ziegler and R. A. Young, Studies of Compounds for Superconductivity, Phys. Rev. **90**, 115 (1953).

[47] J. E. Schirber, D. L. Overmyer, B. Morosin, E. L. Venturini, R. Baughman, D. Emin, H. Klesnar, and T. Aselage, Pressure dependence of the superconducting transition temperature in single-crystal NbB$_x$ (x near 2) with $T_c$=9.4 K, Phys. Rev. B **45**, 10787 (1992).

[48] H. Kotegawa, K. Ishida, Y. Kitaoka, T. Muranaka, N. Nakagawa, H. Takagiwa, and J. Akimitsu, Evidence for strong-coupling s-wave superconductivity in MgB2: 11B-NMR study of MgB2 and the related materials, Physica C: Superconductivity **378-381**, 25 (2002).

[49] D. Kaczorowski, A. Zaleski, O. Żogał, and J. Klamut, Incipient superconductivity in TaB2, arXiv preprint cond-mat/0103571 (2001).

[50] S. Sanfilippo, H. Elsinger, M. Núñez-Regueiro, O. Laborde, S. Lefloch, M. Affronte, G. L. Olcese, and A. Palenzona, Superconducting high pressure CaSi2 phase with $T_c$ up to 14 K, Phys. Rev. B **61**, R3800 (2000).

[51] S. Gabani, K. Flachbart, K. Siemensmeyer, and T. Mori, Magnetism and superconductivity of rare earth borides, Journal of Alloys and Compounds **821**, 153201 (2020).

[52] C. Parlak, The physical properties of AlB2-type structures CaGa2 and BaGa2 : An ab-initio study, Physica B: Condensed Matter **576**, 411724 (2020).

[53] F. Bouquet, R. A. Fisher, N. E. Phillips, D. G. Hinks, and J. D. Jorgensen, Specific Heat of Mg11B2: Evidence for a Second Energy Gap, Phys. Rev. Lett. **87** (2001).

[54] K.-P. Bohnen, R. Heid, and B. Renker, Phonon dispersion and electron-phonon coupling in MgB 2 and AlB 2, Phys. Rev. Lett. **86**, 5771 (2001).

[55] G. V. Samsonov and I. M. Vinitskiĭ, *Handbook of refractory compounds* (Springer, 1980).

[56] S. M. Sichkar and V. N. Antonov, Electronic structure, phonon spectra and electron–phonon interaction in ScB2, Low Temperature Physics **39**, 595 (2013).

[57] A. S. Cooper, E. Corenzwit, L. D. Longinotti, B. T. Matthias, and W. H. Zachariasen, Superconductivity: The Transition Temperature Peak Below Four Electrons per Atom, Proceedings of the National Academy of Sciences **67**, 313 (1970).